\documentclass[journal]{IEEEtran}
\usepackage{cite}
\usepackage{amsmath,amssymb,amsfonts}
\usepackage{algorithmic}
\usepackage{graphicx}
\usepackage{subfig}
\usepackage{textcomp}
\usepackage{xcolor}
\usepackage{comment}
\usepackage{hyperref}
\newcommand{\myhash}{\raisebox{\depth}{\#}}

\hypersetup{
    colorlinks=true,
    linkcolor=blue,
    filecolor=magenta,      
    urlcolor=blue,
}

\begin{document}


\title{Super-Resolving Beyond Satellite Hardware Using Realistically Degraded Images}

\author{Jack~White,~\IEEEmembership{Graduate Member,~IEEE,}
        Alex~Codoreanu,
        Ignacio~Zuleta,~\IEEEmembership{Member,~IEEE,}
        Colm~Lynch,
        Giovanni~Marchisio,~\IEEEmembership{Member,~IEEE,}
        Stephen~Petrie and
        Alan~R.~Duffy
        
\thanks{J. White was with the Department of Computer Science and Software Engineering, Swinburne University of Technology, Hawthorn, VIC, AUS.}
\thanks{A. Codoreanu was with the Gravitational Wave Data Centre, Swinburne University of Technology, Hawthorn, VIC, AUS.}
\thanks{A. Duffy was with the Space Technology and Industry Institute, Swinburne University of Technology, Hawthorn, VIC, AUS.}
\thanks{I. Zuleta, C. Lynch and G. Marchisio were with Planet Labs Inc., San Francisco, CA, 94107 USA.}
\thanks{S. Petrie was with the Centre for Transformative Innovation, Swinburne University of Technology, Hawthorn, VIC, AUS.}}


\maketitle

\begin{abstract}
  Modern deep Super-Resolution (SR) networks have established themselves as valuable techniques in image reconstruction and enhancement. However, these networks are normally trained and tested on benchmark image data that lacks the typical image degrading noise present in real images. In this paper, we test the feasibility of using deep SR in real remote sensing payloads by assessing SR performance in reconstructing realistically degraded satellite images. We demonstrate that a state-of-the-art SR technique called Enhanced Deep Super-Resolution Network (EDSR), without domain specific pre-training, can recover encoded pixel data on images with poor ground sampling distance, provided the ground resolved distance is sufficient. However, this recovery varies amongst selected geographical types. Our results indicate that custom training has potential to further improve reconstruction of overhead imagery, and that new satellite hardware should prioritise optical performance over minimising pixel size as deep SR can overcome a lack of the latter but not the former.
\end{abstract}


\IEEEpeerreviewmaketitle

\vspace{-0.5cm}
\section{Introduction}
\label{sec: introduction}

Remote sensing systems are fundamentally bandwidth limited by the hardware specifications of their imaging systems. For satellites this equates to a limitation in the resolution of captured images, given by the Ground Resolving Distance (GRD). Therefore, a highly desirable outcome for remote sensing is to surpass the GRD of a given system to identify greater detail in images without costly hardware upgrades.

One solution is to leverage modern deep Super-Resolution (SR) networks for extracting and enhancing image features beyond their base resolution. SR generally refers to the task of obtaining a High-Resolution (HR) output from a Low-Resolution (LR) input, with networks typically trained on LR/HR pairs. This is a challenge for remote sensing applications where data acquisition is expensive and the sparsity of suitable LR/HR pairs is prominent. Therefore, an inexpensive method for upscaling remote sensing image data is to leverage pre-trained deep SR networks as no further data acquisition is required and the performance is state-of-the-art.  

The SR reconstruction of images is limited by the information within the input image itself. In satellite imaging, this is driven by three factors: quality of the satellite's focus instrument (ie. mirror), physical spacing between the individual cells within the charge-coupled device (CCD), and the available {\it nadir} time. These challenges are not unique to satellites and SR techniques have seen success in fields such as medical imaging \cite{robinson2017new} and security \cite{uiboupin2016facial}. Thus, an investigation of the performance of deep SR networks and their robustness to known degradation effects in remote sensing data is critical to inform the acquisition of data and future training of SR networks in this field.

\begin{figure}[!h]
\vspace{-10pt}
\centering
\includegraphics[scale=0.4]{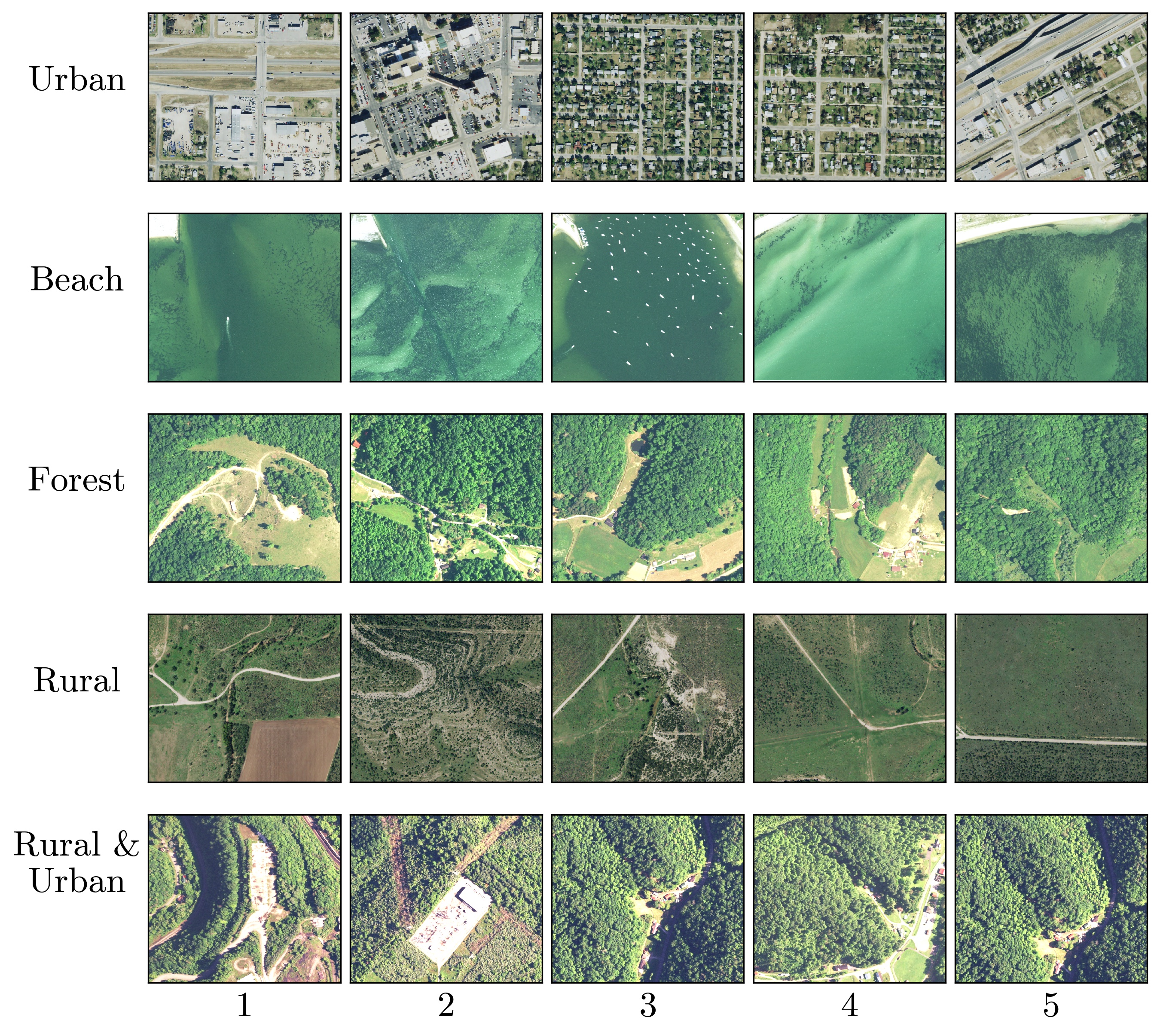}
\caption{Crops from NAIP imagery for each geographical type.}\label{fig:all_terrains_and_crops}
\end{figure}
\vspace{-10pt}

In this paper, we deploy a combined image degradation and SR pipeline to simulate the conditions of a real remote sensing imaging system on a set of sampled National Agricultural Imagery Program (NAIP) $60\text{cm}$ images shown in Figure \ref{fig:all_terrains_and_crops}. We apply realistic degradation to these images and super resolve them using the pre-trained Enhanced Deep Super-Resolution Network (EDSR) \cite{lim2017enhanced}. EDSR was selected for this study as it was the highest performing general, deep SR network at the time of writing and is not domain specifically trained. We use Image Quality Assessment (IQA) metrics to assess the performance of EDSR across a number of realistic degradation parameters. Our assessment is an exploration of the optimal design of remote sensing payloads for SR image reconstruction on satellite image data, with the goal of identifying potential trade-offs between image resolution and cost, thus informing the future construction of remote sensing devices. Finally, we explore a range of terrain image types (see Figure \ref{fig:all_terrains_and_crops}) to assess the versatility and robustness of SR Deep Neural Network (DNN) techniques to the diverse sets of visual features present in Earth Observation (EO) data. 

This paper is structured as follows: Section \ref{sec: background} discusses the pertinent literature and the contribution our paper makes to the field. The methods of our work are then described in Section \ref{sec: method}. In Section \ref{sec: results}, we present our results of the degradation assessment and provide visual example of the super resolved images. We then discuss these results and their implications in Section \ref{sec: discussion} and provide our conclusions in Section \ref{sec: conclusion}.

\begin{figure*}[!h]
\includegraphics[width=\textwidth]{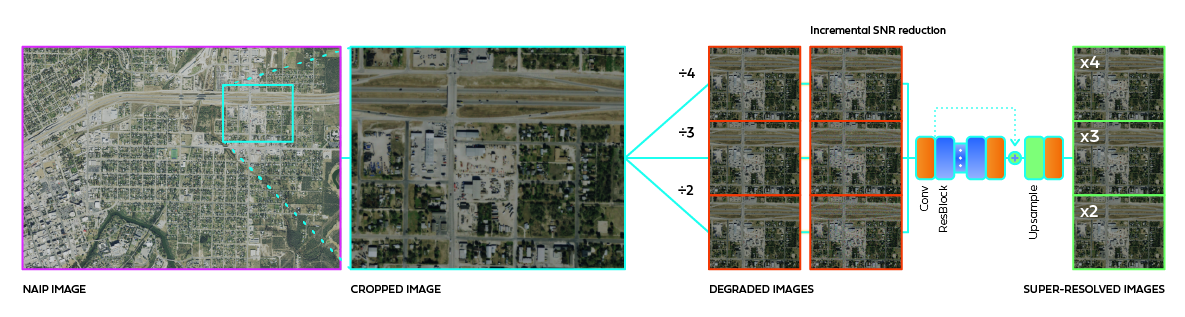}
\caption{Example schematic of a crop from NAIP imagery of the urban geography type with the systematic degradation in GSD, GRD and SNR space. These degraded images are then passed through the EDSR network to super-resolve them back for direct comparison with the original crop. For more detail on the processes in the selection of the imagery see Section~\ref{sec: method}.}
\label{fig:pipeline}
\vspace{-15pt}
\end{figure*}

\section{Background}
\label{sec: background}


Investigations of SR in remote sensing have extensively explored the use of Deep Neural Networks (DNNs). Both Single Image Super Resolution (SISR) \cite{jiang2019edge, tuna2018single, salgueiro2020super, pouliot2018landsat} and Multiple Image Super Resolution (MISR) \cite{molini2019deepsum, deudon2020highres, salvetti2020multi, rifat2020multi} architectures to super-resolve satellite images have surpassed the performance of traditional SR techniques. Research  exploring the nuances of remote sensing training data to enhance the performance of these DNNs is less common. It has been shown however \cite{lanaras2018super} that DNNs trained on a geographically diverse dataset do indeed outperform those trained on less varying data. Other works \cite{zhu2020super} further emphasise the need for training SR DNNs on domain-specific data by augmenting training data using a realistic image degradation pipeline, demonstrating improved performance on real test data over training on non-degraded data.  Multi-task applications, such as SR-to-object detection pipelines, could also greatly benefit from training data that is geographically diverse and includes realistic noise. Examples of deep SR networks that leverage domain specific training include the Deep Distillation Recursive Network (DDRN) \cite{jiang2018deep} and a Coupled Sparse Autoencoder (CSAE) \cite{shao2019remote}. Our paper further explores these data-centric issues in remote sensing by analysing geographical dependencies in image reconstruction.

A key constraint on data-driven SR is the concealing of critical visual information by naturally occurring obscurations such as shadows, cloud coverage and temporal variability in images. EO image data is extremely diverse in terms of the number of present features within heavily complex visual representations. This can include structured urban environments, relatively featureless deserts and oceans, and recurrent forest scenes that vary greatly overall in contrast and organisation. In addition to the data collection cost, the data itself is highly complex and resource-intensive to analyse. 

Deep SR architectures trained on visually diverse datasets such as DIV2K \cite{timofte2017ntire} and RealSR \cite{cai2019toward} have the potential to mitigate some of the needs for domain specific training given the diverse features present in these benchmarks. While not tested on EO image data, the EDSR network outperformed state-of-the-art competitors on the DIV2K dataset. The Very Deep Persistent Memory Network (MemNet) architecture also yielded comparable results with fewer parameters making it more computationally efficient \cite{tai2017memnet}. Finally, the U-shaped Deep Super Resolution network, trained on the RealSR dataset, generates highly accurate reconstructions of complex real images \cite{cai2019ntire}. With the diversity of images in the aforementioned datasets, and generalisability of these trained networks to input image data, there is significant potential benefits for remote sensing applications. That degree of applicability is explored in this paper.

\section{Method}
\label{sec: method}
In this paper, we explore the performance of SR techniques on recovering imagery after realistic, and highly non-linear, degradation steps on original NAIP 60cm imagery. These images are selected by eye to feature a broad geographic range across the USA States of Utah, Kentucky and Massachusetts; qualitatively labeled as Beach, Forest, Rural, Rural with Urban, and Urban (Figure \ref{fig:all_terrains_and_crops}). After identifying a high-resolution image of each type, five crops were selected by hand across each and used as input to the image processing pipeline.

As shown in Figure \ref{fig:pipeline}, the pipeline first systematically degrades each image via a set of realistic degradation parameters. We explore a degrading of Ground Resolved Distance (GRD; effectively optical resolution of the satellite), increasing of Ground Sampling Distance (GSD; essentially decreasing the pixel density of the camera) and multiple Signal-to-Noise Ratio (SNR) levels (representing a change in exposure time). This process is explained in Section~\ref{subsec: degrade}. The second stage of our pipeline then super resolves degraded images using EDSR at each of its upscaling resolutions (x2, x3, x4), as detailed in Section~\ref{subsec: dsr}. Finally, we assess SR recovery performance of the resulting images using two standard image quality metrics SSIM and PSNR. Each metric is normalised to the original input image. These metrics are explained in Section~\ref{subsec: IQA}.

\subsection{Degradation Pipeline} 
\label{subsec: degrade}
The degradation module of our pipeline takes a single image and systematically adds noise to the image according to the input parameters. Our algorithm augments the input image in three parts; blurring through convolution, down-sampling, and Poisson noise augmentation. The parameters of our trade-space model (and associated values used to model optical noise) are the GSD $\left(\text{GSD} = 1.2:0.6:2.4 \right)$, signal-to-noise at $50\%$ of the dynamic range $\left(\text{SNR50} = 10:10:100\right)$ and the GRD $\left(\text{GRD} \approx 1.2:0.35:2.7 \right)$.

To simulate optical degradation, we take the original NAIP selection, $I_{orig}$, and convolve with the Point Spread Function (PSF) of the system. The PSF was generated by the inverse Fourier transform of the Modulation Transfer Function (MTF), simulated by the auto-correlation of the pupil function. An occluded aperture with inner-to-outer diameter radio of $0.4$ was assumed for the pupil. This PSF is scaled with a central MTF padded to match the underlying GRD according to
\begin{equation}
\label{eq: GRD}
    GRD = \frac{1.22 \lambda H}{D} \,,
\end{equation}
where $D = N \lambda H$ has $N$ steps sample around the fiducial height $H = 500 {\rm km}$ and a central wavelength $\lambda = 560 {\rm nm}$. A 2D convolution of the PSF is applied to the three colour channels separately to produce the blurred image
\begin{equation}
\label{eq: Blur}
    I_{blur} = I_{orig} * PSF \,.
\end{equation}
This blurred image is resized to the ratio of the original GSD (set to be $1 {\rm m}$) and sensor \text{GSD$_{sensor}$} to produce an effective sensor image $I_{sensor}$ as an intermediary step. We simulate a varying signal-to-noise by sampling from a Poisson distribution for each pixel, with mean $\mu$ given by the sensor image's pixel intensity and $bit=8$ accuracy for NAIP, set by 
\begin{equation}
\label{eq: mu}
    \mu = I_{sensor} W / 2^{bit}  \,,
\end{equation}
where the equivalent well capacity $W \equiv 2 \times (\text{SNR50})^{2}$. The resulting pixel intensity at each cell is normalised by $W \times 2^{16}$ to give the sampled image $I_{sample}$. Finally, this image is further downsampled to the desired GSD for comparison by resizing with the ratio \text{GSD$_{sensor}$}/\text{GSD$_{product}$}, where \text{GSD$_{sensor}$} is $3 {\rm m}$. This process generates a set of images that capture the spectrum of optical noise over the trade-space.

\subsection{Deep Super Resolution}
\label{subsec: dsr}
For our SR pipeline we elected to use the state-of-the-art, single image, SR network EDSR \cite{lim2017enhanced}, which has a range of upscaling capabilities and highly accurate super resolution performance. This architecture was pre-trained to upscale input images at $\times2$, $\times3$ and $\times4$ their original resolution. EDSR was trained on the DIV2K high-resolution image data set \cite{timofte2017ntire} with the $\times2$ network trained from scratch, and the $\times3$ and $\times4$ further optimised using the $\times2$ as a baseline. We clarify that no training was undertaken in this study as our intention is to explore the applicability of pre-trained networks to remote sensing image data. Furthermore, we only assess EDSR as it marginally outperforms similar networks, and we argue that results from this study will be reflective of other networks of its kind (see \cite{yang2019deep} for contemporary benchmark results). Training procedures are detailed in \cite{lim2017enhanced} and their code implementation (also used in this paper) is at this \href{https://github.com/LimBee/NTIRE2017}{GitHub link}\footnote{https://github.com/LimBee/NTIRE2017}.

The EDSR architecture is a DNN containing residual blocks of similar network layers. Each block contains two convolution layers with a ReLU activation function in between. The model we employed had $32$ layers with each convolution layer containing 64 feature maps. The residual blocks then feed into the upscaling layers of the network that progressively increase the dimensionality of the input image. The number of upscaling layers is dependent on the upscaling factor, with more layers added as the value increases from $\times 2$ to $\times 4$. 

\subsection{Image Quality Assessment (IQA)}
\label{subsec: IQA}
As is customary in the literature, we use Peak Signal-to-Noise Ratio (PSNR) and Structural Similarity Index (SSIM) IQA metrics to quantify the upscaled image quality \cite{wang2020deep}. Further IQA metrics were tested in this study (e.g. ERGAS, UIQI) but were found to have too little dispersion in results to draw meaningful conclusions from, compared to the richer SSIM and PSNR metrics. The PSNR measures the reconstruction quality of data when passed through a lossy transformation. We define the PSNR as a value scaled by the mean-squared error (MSE) of the difference in pixel intensities between the ground-truth image $I$ and reconstruction $\hat{I}$ both with $N$ pixels
\begin{equation}
\label{eq: PSNR}
    PSNR = 10\times\log_{10}\left(\frac{L^{2}}{MSE} \right)\,,
\end{equation}
with $L$ the maximum pixel intensity value and MSE given by
\begin{equation}
\label{eq: MSE}
    MSE = \frac{1}{N}\sum^{N}_{i=1}{\left(I(i) - \hat{I}(i)\right)^{2}}\,.
\end{equation}
\noindent
The SSIM incorporates the intensity, contrast and variance of each pixel and approximates the human perceived image quality within a 0 to 1 range. It can be represented as
\begin{equation}
\label{eq: SSIM}
\text{SSIM}(I, \hat{I}) = \frac{\left(2\mu_{I}\mu_{\hat{I}} + C_{1}\right)\left(\sigma_{I \hat{I}} + C_{2} \right)}{\left(\mu_{I}^{2} + \mu_{\hat{I}}^{2} + C_{1}\right)\left(\sigma_{I}^{2} + \sigma_{\hat{I}}^{2} + C_{2} \right)} \,,
\end{equation}
\noindent 
where $\mu$ and $\sigma$ are mean and standard deviation of the image intensity for reference $(I)$ and sample $(\hat{I})$ images, and $C_{1}$ and $C_{2}$ are constants that stabilise the metric calculation. As it depends on contrast, this metric is sensitive to cutoffs in the high frequency range of an image's power spectrum where contours of fine grain features most likely exist.

\begin{figure}
\centering
\includegraphics[scale=.35]{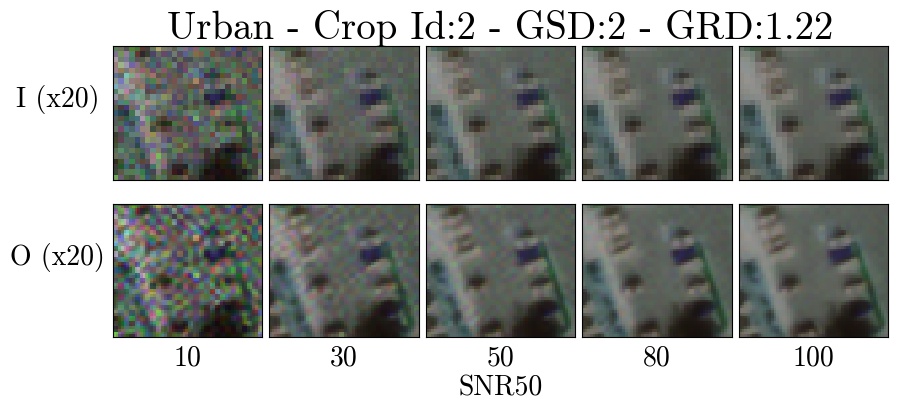}\\
\includegraphics[scale=.35]{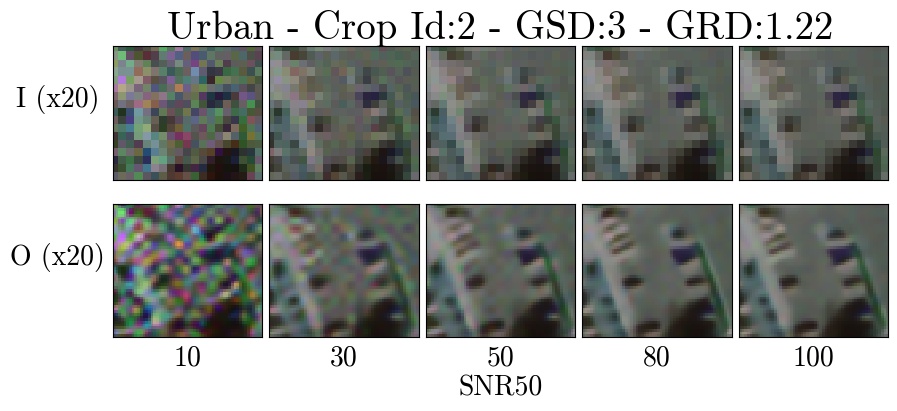}\\
\includegraphics[scale=.35]{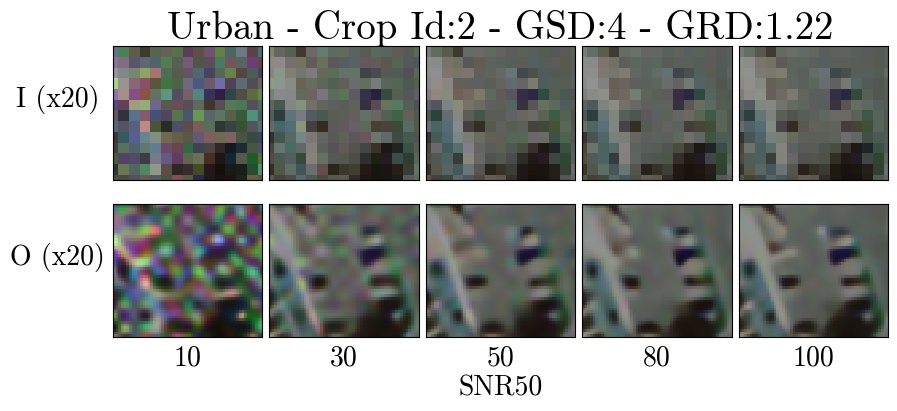}
\caption{Example performance of the EDSR pipeline for the urban geographical type, zoomed in a factor 20 on a car park. Degraded images (I) are shown for a given GSD/SNR trade-space together with the associated SR pipeline output (O) for x2, x3 and x4 degradation (top to bottom, respectively).}
\label{fig:io_comp}
\vspace{-12pt}
\end{figure}

\begin{figure}[!h]
\includegraphics[width=.495\linewidth]{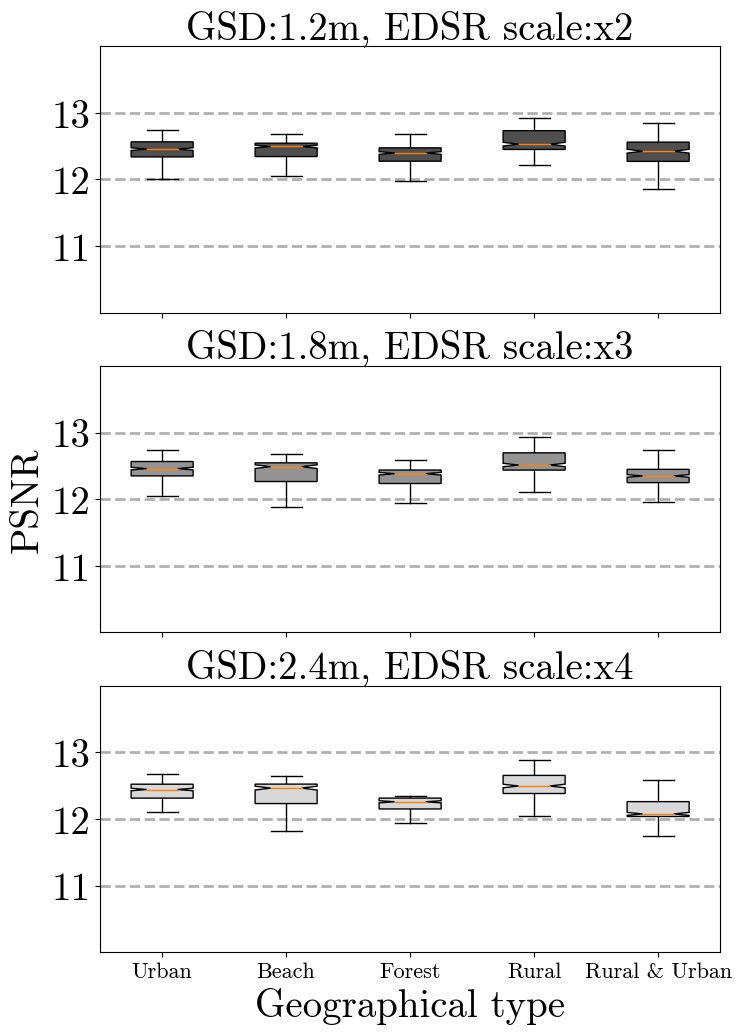}
\includegraphics[width=.495\linewidth]{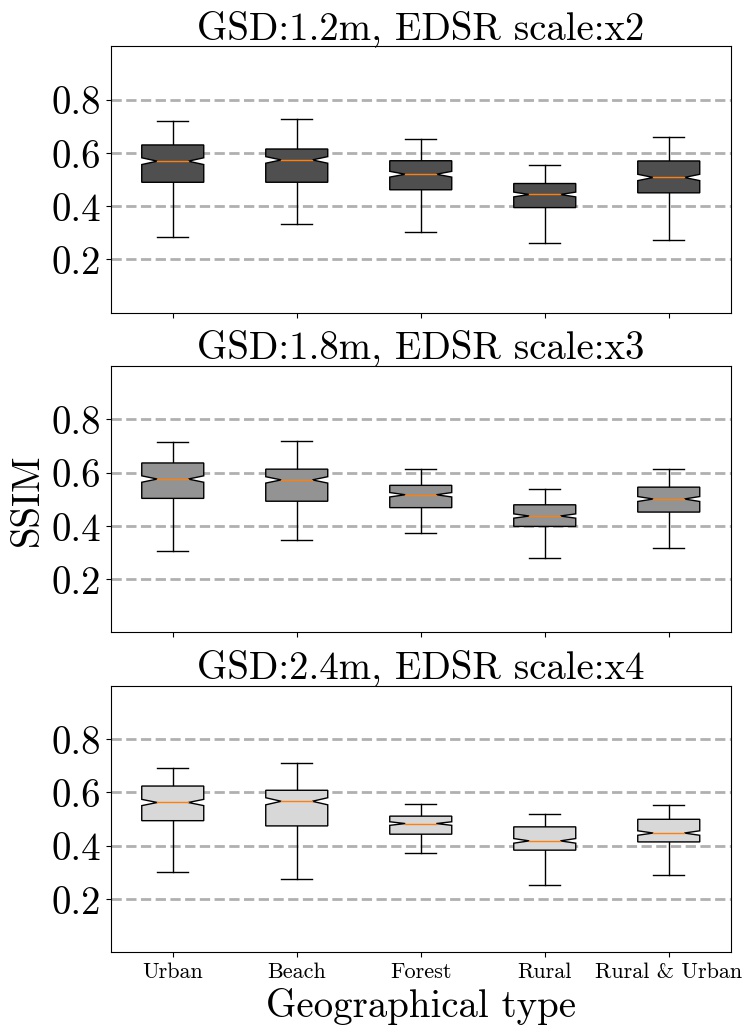}
\caption{A box plot demonstrating the recovery performance of the SR technique for various geographical types as assessed by the PSNR (SSIM) metric in the left (right) column, with increasingly degraded GSD factors from top to bottom. The individual box plots are over the full $\text{SNR50}$ range, with median as the orange line and quartiles indicated by the box (and min/max values the whiskers). The negligible change of PSNR with geography, GSD degradation or even $\text{SNR50}$ range is suggestive that it is too simplistic an assessment metric. The SSIM has a more complex behaviour as described in the main text, and is explored in greater detail in Figure \ref{fig:ssim_multi_geo}.}
\label{fig:multiplot}
\vspace{-15pt}
\end{figure}
\noindent
\section{Results}
\label{sec: results}
\noindent
We investigated the PSNR \& SSIM distributions of the super resolved images relative to the original cropped images. These original images were first passed through the {\it degradation pipeline} described in Section \ref{subsec: degrade} after which they were super resolved as described in Section \ref{subsec: dsr}. 

An overview of the degraded crops, and the output from the SR pipeline, is demonstrated in Figure \ref{fig:io_comp}. 
The dramatic visible recovery of cars from images degraded even by a factor 4 in GSD (at high SNR; bottom right) is a qualitative demonstration of the EDSR success of information recovery. To see a gallery of similarly degraded images for each of the other terrain types and crop numbers refer to this \href{https://smpetrie.github.io/superres/}{GitHub link}.

We assess the performance by grouping the output of each individual scaling factor used in the SR pipeline (x2, x3, x4) and geographical subset. We consider all {\it SNR50} values associated within each subset and create a box-plot over these values, for each metric considered. The PSNR (SSIM) distribution can be seen in the left (right) column of Figure \ref{fig:multiplot}. We note that the sample size across geographical types is relatively limited in this proof of concept study, but the trends within a geographical type over the trade-space is more statistically rigorous and hence the results presented here are for qualitative rather than quantitative consideration. 

\subsection{PSNR distribution}
\label{sec:psnr_distribution}

As can be seen in the left column of Figure \ref{fig:multiplot}, the PSNR metric has a near constant relation for each geographical and GSD subset with a median value between 12 and 13 for all subsets.
This indicates that the EDSR network is invariant to both parameters and does not depend on the SNR50 value of the input image. 
However, PSNR only depends on relative pixel intensity (see eq. \ref{eq: PSNR}) and does not provide information on the likely information loss which should not only depend on signal-to-noise but also on the relationship between the GSD, the {\it on the ground} distance between pixels, and the GRD, the minimum physical size that can be resolved by the satellite. 

As can be seen by the left column of Figure \ref{fig:multiplot} there is almost no change between the PSNR metric for different geographical types, GSD scaling or even a particularly significant spread in the PSNR values across the range of $\text{SNR50}$ considered. As a result of the negligible change in the PSNR across this trade-space we argue it makes for a poor IQA for our investigation and we do not consider it further in our analysis. 

\subsection{SSIM distribution}
\label{sec:ssim_distribution}

Unlike the near-invariant PSNR distribution in section \ref{sec:psnr_distribution}, the computed SSIM values span almost half of the possible range as shown in the right column of Figure \ref{fig:multiplot}. The median values associated with each geographical label decrease modestly with increasing GSD values, but with a large spread within each sample. This suggests that the EDSR network struggles to reconstruct features which have been blurred over each other, as the physical size of the blurring kernel increases during the degradation process discussed in section \ref{subsec: degrade}.

\begin{figure}
\vspace{-10pt}
\includegraphics[width=6.75cm]{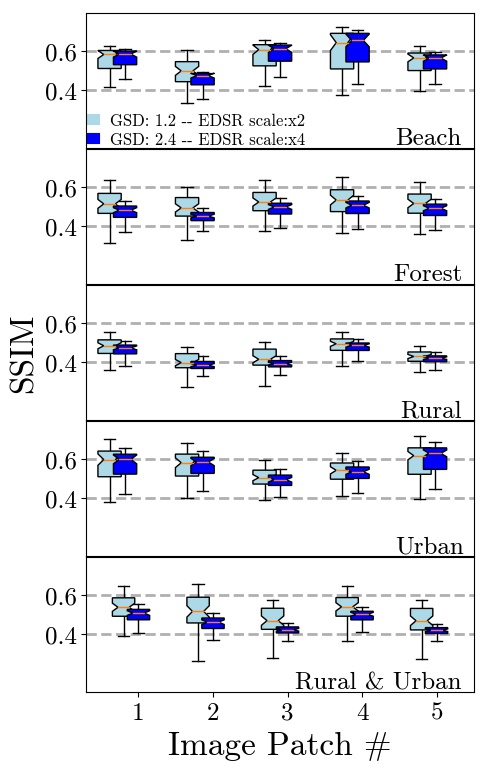}
\vspace{-6pt}
\caption{Similar to the SSIM box plot of Figure \ref{fig:multiplot}, but now exploring the individual patches within a given geographical type. There are small but essentially limited differences of the performance of the SR within a given geographical region.}
\label{fig:ssim_multi_geo}
\vspace{-10pt}
\end{figure}

\begin{figure*}[!h]
\centering
\includegraphics[width=.32\textwidth]{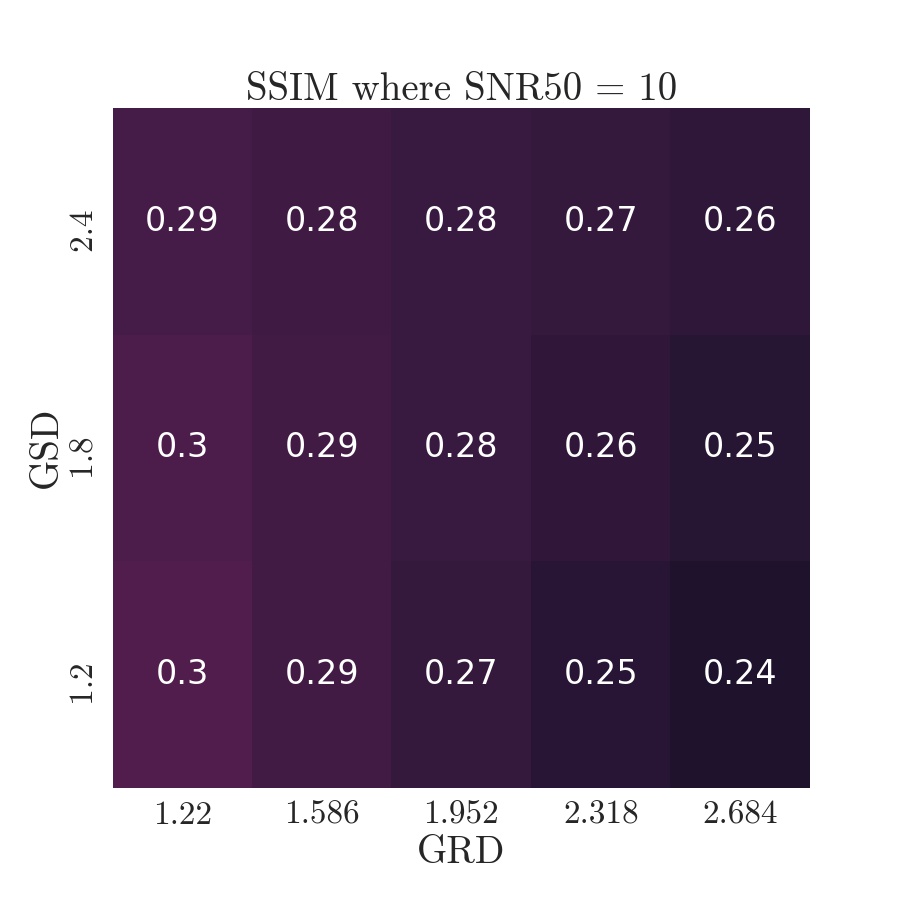} \includegraphics[width=.32\textwidth]{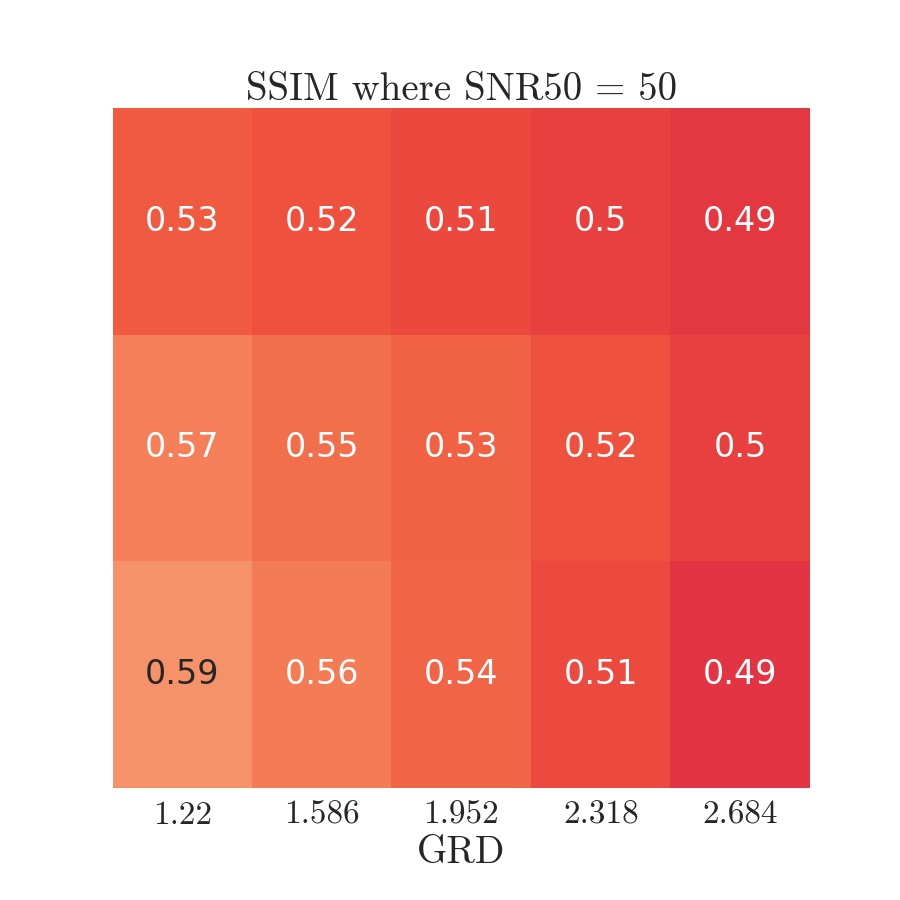} \includegraphics[width=.32\textwidth]{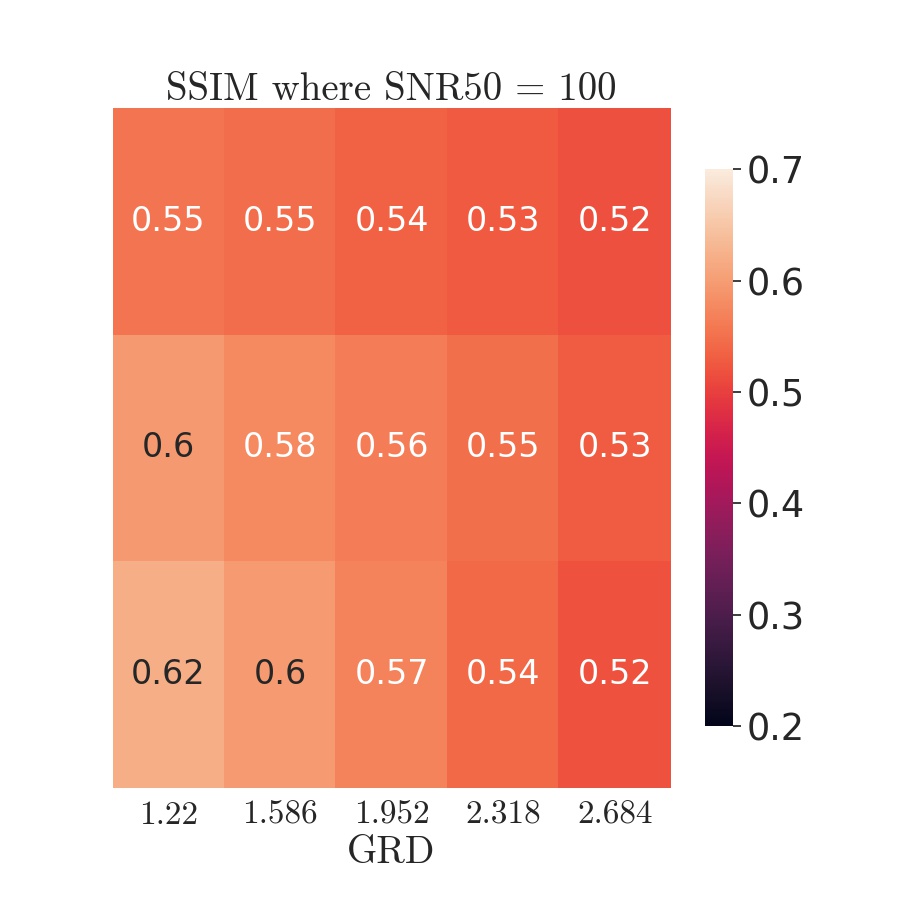}
\vspace{-10pt}
\caption{SSIM heat-maps exploring the GSD and GRD recovery metric for all geographical types, from SNR50 of 10, 50 and 100 in the left, middle and right panels respectively. We observe little performance penalty with increasing the GSD for a given GRD at fixed SNR50, suggesting optics trumps pixels in this trade-space. The SNR50 drives a doubling of SSIM when increased from $\text{SNR50} = 10$ to $50$, but limited improvement thereafter to $\text{SNR50} = 100$, suggesting time on target can be reduced for a satellite without reduced performance when applying SR techniques.}
\label{fig:ssim_heatmaps}
\vspace{-13pt}
\end{figure*}

In order to understand the SSIM distribution associated with each geographical label, we look at the individual performance of each of the five images within each subset. The goal is to understand if the large variance observed in the right column of Figure \ref{fig:multiplot} is consistent across each subset. 

We compare the SSIM distribution for the range of $\text{SNR50}$ values for a given geography and Crop ID (i.e. the same cropped zoom in) in Figure \ref{fig:ssim_multi_geo} with the left (right) column indicating GSD scaling by x2 (x4). 
There is effectively no resulting performance difference for a given Crop ID, i.e. a sample within region type. 
Thus for each geographical subset the SSIM distribution seen in Figure \ref{fig:multiplot} is not driven by a selection effect on how we constructed/sampled each cropped subset image. However, we note the average value across each geographical subset/panel in Figure \ref{fig:ssim_multi_geo} shows some variance, suggesting the network is biased toward the features present in the Beach subset. In particular, Beach Patch \myhash 4 shows an almost 50\% improvement relative to Rural \& Urban Patch \myhash 5.

We next explore the relative performance of GSD and GRD as a function of SNR50. In Figure \ref{fig:ssim_heatmaps} we see that, for a given SNR, the recovery performance of the SR technique as measured by the SSIM metric depends mostly on the GRD. The SNR50 also shows diminishing returns; i.e. comparing $\text{SNR50} = 10$ to $50$ (left to middle panels) the SSIM approximately doubles for all GSD/GRD combinations. However, thereafter to $\text{SNR50} = 100$ sees little increase in the SSIM, suggesting that for these images the signal-to-noise ratio could effectively be halved with negligible reduction in satellite performance when combined with super resolving techniques. In addition, a custom trained EDSR network could also increase the SSIM location for this plateau or even further increase the reconstruction quality of images with SNR50$>$50.

\section{Discussion}
\label{sec: discussion}
Our results show that the performance of current deep super resolution techniques on remote sensing data is weakly dependent on the geographical type of data. Figure \ref{fig:ssim_multi_geo} indicates that particular geographical types achieve higher SSIM values on average when images with more distinct features are present. The SSIM show well defined trends where the peak mean for the highest SNR50 values are limited by the geographical type of the image. Given that the largest SNR50 values correspond to the least noisy images, this suggests that geotype specific training could enhance the capabilities of SR networks on remote sensing image data. We remind the reader this study intentionally explored the use of EDSR networks pre-trained on DIV2K image data. It is, therefore, not unreasonable to assume network performance in this study could be improved through training on domain specific satellite imagery. 

In conjunction with this procedure, a realistic training data set must be augmented to contain realistic noise as a means of increasing the robustness of the network to typical remote sensing optical noise. Using the simple model outlined in Section \ref{subsec: degrade}, training data can be generated with varied noise by randomising the parameters over a realistic trade-space.

As presented in section \ref{sec:psnr_distribution}, the PSNR shows little trend with variations in the SNR50 and image re-scaling, suggesting it is ill-suited as an IQA for this work. The overall statistical strength of these conclusions will be explored in future work with larger sample sizes, but as a proof of concept these findings are certainly suggestive that SSIM is a more suitable IQA metric for satellite imagery than PSNR. In addition, a future robust metric must also account for the actual retrieval of objects from the reconstructed vs. training data set.

\section{Conclusion}
\label{sec: conclusion}
In this paper, we established the feasibility of using pre-trained deep super resolution networks to enhance remote sensing images that have undergone realistic image degradation, and shown that a pre-trained network can compensate for certain satellite hardware limitations. In particular we explored the performance of the Enhanced Deep Super Resolution (EDSR) network in super resolving realistically degraded images across a degradation trade-space that included the Ground Sampling Distance (essentially the distance on ground that each camera pixel `sees'), the Ground Resolving Distance (the smallest feature on the ground that the satellite can resolve) and the overall Signal-to-Noise of the image (representing the performance of the camera / exposure time).

Assessing the network with two standard Image Quality Assessment techniques, we found the oft-used PSNR is insensitive to modifications and unsuited to this kind of study, unlike the SSIM which we utilised throughout this paper. SSIM reveals super resolution is well suited to recover a degradation of Ground Sampling Distance, provided the Ground Resolving Distance is higher; more colloquially, optics trump pixels. 

Our findings have important ramifications for satellite payload and operational considerations. The network had a plateauing response to the Signal-to-Noise, with great improvement in image quality when SNR50 was increased from 10 to 50 but little improvement for further signal increase. Thus it may be feasible to dramatically reduce exposure times if a super resolution network is employed to remove added noise. Finally, we have shown that the geographical content of images will influence the reconstruction quality from the super resolution network, suggesting that bespoke training may yet yield even greater performance in remote sensing images. 

\section*{Acknowledgment}
The authors acknowledge the support, data contributions and availability of satellite data from Planet Labs Inc, and thanks the United States Department of Agriculture for making NAIP imagery freely available. Support for the project has been provided by Swinburne University of Technology's Data Science Research Institute. The work in this paper made use of the OzSTAR national HPC facility. OzSTAR is funded by Swinburne University of Technology and the National Collaborative Research Infrastructure Strategy.

\bibliographystyle{IEEEtran}
\bibliography{IEEE_TGRS/IEEE_TGRS_refs}

\end{document}